\documentclass[preprint,eqsecnum,aps,emlines2,bezier,showpacs,draft]{revtex4}
\def\eps{\varepsilon}
\def\partt{\mbox{\boldmath $\partial$}}

\def\dhline#1#2#3#4#5#6{  {\countdef\nnn=255 \dimendef\llx=0   \dimendef\lly=1 \dimendef\dx=2   \dimendef\dy=3\llx=#1\unitlength  \lly=#2\unitlength \dx=#4\unitlength\dy=#5\unitlength  \nnn=#6 \divide\nnn by 2 \advance\dx by-\llx\advance\dy by-\lly \div\nnn \div4 \lline \adv \multiply\dx by2 \multiply\dy by2 \loop \adv \ifnum\nnn>1 \lline \adv \advance\nnn by-1 \repeat \div2 \lline }}

\def\div#1{ \divide\dx by#1  \divide\dy by#1 }
\def\adv{ \advance\llx by\dx \advance\lly by\dy }

\def\lline{ {  \divide\llx by\unitlength \divide\lly by\unitlength \divide\dx by\unitlength \divide\dy by\unitlength \advance\dx by\llx \advance\dy by\lly\emline{\number\llx}{\number\lly}{}{\number\dx}{\number\dy}{}}}

\newcommand{\D}{\displaystyle}

\begin{document}


\title{Large-order asymptotes of the quantum-field expansions for Kraichnan model of passive scalar advection }

\author{Andreanov A.Yu., Komarova M.V., Nalimov M.Yu.}

\affiliation{Department of Statistical Physics, St.- Petersburg State University, Uljanovskaja 1, St.- Petersburg, Petrodvorez, 198504  Russia}

\date{\today}


\begin{abstract}
A simple model of a passive scalar quantity advected by a Gaussian non-solenoidal (``compressible'') velocity field is considered. Large-order asymptotes of quantum-field expansions are investigated by instanton approach. The existence of finite convergence radius of the series is proved, a position and a type of the corresponding singularity of the series in a regularization parameter $\eps$ are determined. Anomalous exponents of the main contributions to the structural functions are resumed using new information about the series convergence and two known orders of the $\eps$ expansion.
\end{abstract}


\maketitle

\section{Introduction}
\label{sec:1}

The renormalization group can be considered as a most powerful method of investigation of the critical and scaling behavior. It produces results in a form of some, usually asymptotic, expansions and only a few first terms are known analytically. Different resummation techniques are used to obtain reliable results \cite{Zinn}. The large-order asymptotic information of perturbation series for field-theoretic models is the base of critical exponents and scaling functions series resummation \cite{Zinn}. This information has been obtained with the aid of instanton approach at the example of the static $\phi^4$ model and applied to the resummation problem \cite{Lipatov,ZinnBr,TMF}.

Unlike the static models with 5-6 terms of a regular expansion for critical indexes known the perturbation series in the dynamical ones are calculated up to the third order only \cite{Dominicis,Antonov,AdzhAnt}. Nevertheless the resummation problem arises here also and demands an information about the large-order asymptotic in dynamic field theories. The aim of this paper is to develop the instanton approach for large-order asymptotic investigation in dynamics at the example of a simple dynamical model constructed from Langevin equation in the framework of the MSR formalism \cite{MSR}.

A model of the passive advection of a scalar admixture by a Gaussian short-correlated velocity field, introduced by Obukhov \cite{Obukhov} and Kraichnan \cite{Kraichnan} attracts now a considerable interest. Some first structural and response functions in this model demonstrate an anomalous scaling behavior and the corresponding anomalous exponents can be calculated explicitly as within regular expansions in different small parameters as using numerical simulations. Thus this model provides a good testing ground for various concepts and methods of the turbulence theory like closure approximations, refined similarity relations, Monte-Carlo simulations and renormalization group investigation.

The advection of a passive scalar field $\varphi({\bf x},t)$ is described by a stochastic equation:

\begin{equation}
\label{main}
{\partial}_t\varphi - {\nu}_0\Delta\varphi + g{{\partt}}_i({\bf v}_i\varphi) = \xi({\bf x},t),
\end{equation}

\noindent where ${\partial}_t=\partial/\partial t, {\partt}_i=\partial/\partial {\bf x}_i, {\nu}_0$ is a molecular diffusivity coefficient, $\Delta$ is Laplace operator with respect to ${\bf x}$, $\xi$ is an artificial Gaussian scalar noise with zero mean and given correlator $D_\xi$, ${\bf v}({\bf x},t)$ is a velocity field, $ g $ is a coupling constant; the sum in repeating indices here and henceforth are implied. Besides in the Kraichnan model the field ${\bf v}({\bf x},t)$ obeys a Gaussian distribution with zero average and the following correlator \cite{Adzhemyan}

\begin{eqnarray}
\label{cor1}
{\bf D}_{v}^{ij}({\bf x}-{\bf x}^\prime)\equiv \langle {\bf v}_i({\bf x},t){\bf v}_j({\bf x}^{\prime},t^{\prime})\rangle = D_{0}\delta(t - t^{\prime})\int\frac{d {\bf k}}{{(2\pi)}^d}\D\frac{P_{ij}^{\perp}({\bf k}) + \alpha P_{ij}^{\parallel}({\bf k})}{{({\bf k}^2 + m^2)}^{d/2 + \eps}}e^{i{\bf k}({\bf x} - {\bf x}')}.
\end{eqnarray}

\noindent where $P^{\perp}({\bf k}), P^{\parallel}({\bf k})$ are projectors with respect to vector ${\bf k}$ direction, $d$ is a dimension of the space, $\eps$ is the regular expansion parameter and an ultraviolet (UV) regularizator of the model just as $m$ is an infrared one; $\alpha$ is a coefficient representing non-solenoidal modes contribution, $D_{0}$ is an arbitrary amplitude regular in $\eps$. The tensor indices of correlator will be omitted and implied henceforth.

Using MSR formalism one can formulate the Kraichnan model as a field theory with the following renormalized action \cite{Adzhemyan}

\begin{eqnarray*}
S_{R} = \frac{\varphi' D_{\xi}\varphi' }{2} + \varphi' \Bigg( -\partial_{t}\varphi - g\partt_i({\bf v}\varphi) + \nu Z_{\nu}\Delta\varphi \Bigg ) - \frac{{\bf v}_i{\bf D}_{v}^{-1}{\bf v}_j}{2},
\end{eqnarray*}
integration in field arguments is implied. The renormalization constant $Z_{\nu}$ is determined by the requirement that the 1-irreducible part of the function $\langle\varphi'\varphi\rangle$ expressed in renormalized variables is UV finite (i.e., is finite at $\eps\to0$). This requirement remains for $Z_{\nu}$ an UV finite arbitrariness; the latter is fixed by the choice of a renormalization scheme. The minimal substractions (MS) scheme show it's worth as the most useful for calculations. In this scheme all renormalization constants have the form ``1 + only poles in $\eps$.''  The 1-irreducible part of the function $\langle\varphi'\varphi\rangle$ of Kraichnan model is known exactly. It gives

\begin{equation}
\label{Z}
Z_{\nu}= 1 - u\, D_{0}\Bigg|_{\eps=0}\frac{S_{d}}{(2\pi)^{d}} \frac{d-1+\alpha}{2d\eps},
\end{equation}

\noindent where $ u\equiv g^{2} $, $S_d\equiv 2\pi^{d/2}/\Gamma (d/2)$ is the surface area of the unit sphere in $d$-dimensional space. Note that the result (\ref{Z}) is exact, i.e., there are no corrections of the order $u^{2}, u^{3}$, and so on \cite{Adzhemyan}.

It was stated recently \cite{Chertkov,Balkovsky} the absence of an instanton within the MSR approach in Kraichnan model and it is the Lagrangian variables that can be used at the steepest-descent investigation of the structural functions $<[\varphi(x_{1},t_{1})-\varphi(x_{2},t_{2})]^{n}>$ in the large $ n $ limit. We deal with another problem here, namely the large-order asymptotic investigation for the perturbation series of the anomalous dimensions at arbitrary $n$. Nevertheless we also introduce the Lagrangian variables following \cite{Chertkov,Balkovsky} to construct an instanton.

Let us remind the main features of the Lagrangian variables approach. The Green function G({\bf x},s;{\bf y},t) corresponding to the Eq. (\ref{main}) obviously satisfies

\begin{eqnarray*}
\label{lv-main-green}
({\partial}_s-\nu _{0}\triangle)G({\bf x},s;{\bf y},t) + g{\bf \partt_i }({\bf v_i}({\bf x},s)G({\bf x},s;{\bf y},t)) = \delta({\bf x} - {\bf y})\delta(s - t).
\end{eqnarray*}

We can eliminate the $\delta$-functions in the right-side by taking the following initial object

\begin{equation}
\label{1.5a}
G({\bf x},s;{\bf y},t)=\theta(s - t)P({\bf x}, s;{\bf y},t).
\end{equation}
This yields
\begin{eqnarray*}
\label{lv-main-fokker}
({\partial}_s - \nu _{0}\triangle_{\bf x})P({\bf x},s;{\bf y},t) + g{\bf \partial }({\bf v}({\bf x},s)P({\bf x},s;{\bf y},t)) = 0\\
P({\bf x}, s = t;{\bf y},t)\equiv{\delta}^d({\bf x} - {\bf y}).
\end{eqnarray*}

The last equation may be considered as a Fokker-Plank equation of some different model with $P({\bf x},s)$ being a simultaneous distribution function. The term to term comparison allows us to reconstruct this model and to write down the solution $P({\bf x},s;{\bf y},t)$ explicitly:

\begin{eqnarray}
\label{lv-p-dist-func}
P({\bf x},s;{\bf y},t)=\langle\delta^d({\bf x}-{\bf X}(s)){\rangle}_{\zeta}\\
\label{lv-add-model}
{\partial}_s{\bf X}(s)=-g{\bf v}({\bf X}(s),s)+{\bf \zeta}(s)\\
\nonumber {\bf X}|_{s=t}={\bf y},\quad D_{\zeta}=2\nu_{0},
\end{eqnarray}

\noindent $<\ldots>_{\zeta}$ denotes an average in the auxiliary random force $ \zeta $ with Gaussian distribution ($D_{\zeta}$ introduced is a corresponding correlator). Equations (\ref{lv-p-dist-func},\ref{lv-add-model}) describe particle motion in a random force field $\zeta$. We see that the analysis can be reduced to the study of some interacting particles moving in a random medium as all the quantities in Kraichnan model can be expressed via Green function $C({\bf x},s;{\bf y},t)$ in the random field ${\bf v}$. The medium is inhomogeneous: the Eq.(\ref{lv-add-model}) would be exactly the ordinary Brownian motion equation if there wasn't a term containing the velocity field that depends explicitly on the particle position.

Let's insert in the Eq.(\ref{lv-p-dist-func}) an additional path integration over an auxiliary field ${\bf c}(s)$ having in integrand Dirac ${\delta^F}$ - function with respect to the field variable ${\bf c}$:

\begin{eqnarray}
\label{1.7}
P({\bf x},s;{\bf y},t)=\langle\int D{\bf c}{\delta^F}({\bf c}(s)-{\bf X}(s)) \delta^d({\bf c}(s)-{\bf x}){\rangle}_{\zeta}
\end{eqnarray}

\noindent and rewrite ${\delta^F}$ - function in the form including a functional determinant:

\begin{eqnarray}
\nonumber {\delta^F}({\bf c}(s)-{\bf X}(s))=\det\Bigg({\partial}_s+g\frac{\delta {\bf v}} {\delta {\bf c}}\Bigg){\delta^F}({\partial}_s{\bf c}+g{\bf v}({\bf c},s)-{\bf\zeta}).
\end{eqnarray}

\noindent Then the $d$ - dimensional $\delta^d$ - function in Eq. (\ref{1.7}) can be transformed to the boundary conditions for ${\bf c}(s)$:

\begin{eqnarray}
\nonumber P({\bf x},s;{\bf y},t) = \langle\int\limits_{{\bf c}(t)={\bf y}} ^{{\bf c}(s)={\bf x}}D{\bf c}\det\Bigg({\partial}_s + g\frac{\delta {\bf v}}{\delta {\bf c}}\Bigg){\delta}^F({\partial}_sc + g{\bf v}({\bf c},s) - {\bf\zeta}){\rangle}_{\zeta}.
\end{eqnarray}

\noindent Converting $\delta^F$ - function to a Fourier-like path integral form we get

\begin{eqnarray}
\label{1.8}
P({\bf x},s;{\bf y},t) = M\int\limits_{{\bf c}(t)={\bf y}}^{{\bf c}(s)={\bf x}}D{\bf c}D{\bf c} ^{\prime}\exp(-\nu {\bf c}^{\prime 2} + i{\bf c}^{\prime}\dot {\bf c} + ig{\bf c}^{\prime}{\bf v}({\bf c},\tau))\det\Bigg({\partial}_s + g\frac {\delta {\bf v}}{\delta {\bf c}}\Bigg),
\end{eqnarray}

\noindent the sums in the vector indexes of ${\bf c}(\tau)$, ${\bf c} ^{\prime}(\tau)$, ${\bf v}({\bf c},\tau)$ fields and integration in field arguments are implied. Henceforth $\dot{\bf c}$ denotes $\partial{\bf c}(\tau)/\partial\tau, M$ is a normalization factor appeared due to a functional determinant of the last $\delta^F$ - function transformation. The field theory obtained coincides exactly with the standard one. This can be proved by comparison of the diagrammatic expansions after the proper regularization of the functional determinants arisen. Namely $\det({\partial}_s +g{\delta {\bf v}}/{\delta {\bf c}})$  can be considered as a constant independent on ${\bf v}$, ${\bf c}$ fields,  like in a usual MSR formalism \cite{Adzhemyan} and it can be included in $M$ factor. Thereupon $M$ is determined by the free theory (at $g=0$) and can be restored by comparison with it:

\begin{eqnarray*}
\label{p2-free-Green}
\nonumber\langle{\varphi}({\bf x},t_0){\varphi}^{\prime}({\bf y},t)\rangle\Bigg|_{g=0} = M\int\limits_{{\bf c}(t)={\bf y}}^{{\bf c}(t_0)={\bf x}}D{\bf c}D{\bf c}' \exp\Bigg(-\nu {\bf c}'^2 + {i}{\bf c}'\dot{\bf c}\Bigg) =\\
= \frac{1}{{(4\pi\nu T)}^{d/2}}\exp\left(-\frac{{({\bf x} - {\bf y})}^2}{4\nu T}\right),\qquad T\equiv t_{0}-t.
\end{eqnarray*}

Usually the perturbation expansions series in the quantum-field theories turn out to be asymptotic ones with zero radius of convergence \cite{Zinn}. The situation is different in Kraichnan model: the exact result for $\varphi^2$ composite operator shows that these series have a finite radius of convergence \cite{Adzhemyan}. In this work we prove the following $N$-th term of the quantum-field expansions of Kraichnan model large $N$ limit behavior

\begin{eqnarray}
\label{nomer} \gamma^{(N)}\sim \frac{N^b}{a^N},
\end{eqnarray}
where $b$ characterizes the type of singularity and $a$ corresponds to the radius of convergence. Our aim is to calculate $a$ and $b$ coefficients for anomalous dimensions of composite operators $\varphi^n$. The large-order information completes the perturbative one and allows us to extract singularities from perturbation series and to improve the convergence of the quantum-field expansions.

The article \cite{Chertkov} results in an interesting statement about the saturation of the scaling dimensions $ \Delta _{n} $ of the $\varphi ^{n}$ operators. Namely they are predicted to tend to the finite value at $ n\to \infty $. One of our purposes is to verify this statement for the compressible velocity case on the basis of the results of the resummation for two orders of $\eps$ expansion.

The paper is organized as follows. In Section II the method of renormalization constants calculation is discussed. The general stationarity equations in field variables are obtained in Section III. The solution of these equations is described in Section IV. In Section V we deal with the stationarity equations in coupling constant and coordinate arguments. The integration over the scale parameter is described in Section VI. Section VII is devoted to a reexpansion of the anomalous dimensions of the composite operators $\varphi^n$. The summary is written in Section "Conclusions".

\section{Large-order investigation of renormalization constants}

The quantity of interest for Kraichnan model is, in particular, the infrared behavior of the single-time structural functions

\begin{eqnarray*}
\label{struc}
\langle[\varphi(t,{\bf x})-\varphi(t,{\bf x'})]^n\rangle, \quad r\equiv|{\bf x}-{\bf x'}|.
\end{eqnarray*}

\noindent It is determined by anomalous dimensions $\gamma_{\varphi^n}$ of the composite operators $\varphi ^{n}$, the former have been calculated up to two orders of $ \eps $ expansion \cite{Adzhemyan}.

Let us consider the response functions $\int d{\bf x}_0 dt_0\langle\varphi^n({\bf x}_0,t_0)\varphi' ({\bf x} _1,t_1)\ldots\varphi' ({\bf x}_n,t_n)\rangle$ represented in a form

\begin{equation}
\label{lag}
\int d{\bf x}_0 dt_0\int D{\bf v} G({\bf x}_0,t_0,{\bf x}_1,t_1)\ldots G({\bf x}_0,t_0,{\bf x}_n,t_n).
\end{equation}

Large-order investigation of this expression as a series in $g$ is very difficult because of the presence in the Eq.(\ref{lag}) additional parameters such as $m, {\bf x}_i, t_i, \eps$ that can constitute different combinations comparable with the large parameter $N$ of the steepest-descent method. The same problem for the classical $\varphi^4$ static model was described in \cite{TMF}. Fortunately, as we are interested in the anomalous dimensions $\gamma_{\varphi^n}$ we can limit ourself by the consideration of renormalization constants $Z_{\varphi^n}$ of the composite operators $\varphi^n$:

\begin{eqnarray}
\label{2.11}
\langle{\varphi}^n\varphi' \ldots\varphi' {\rangle}^R = Z_{{\varphi}^n}^{-1}\langle{\varphi}^n\varphi' \ldots\varphi' \rangle.
\end{eqnarray}

\noindent Constants $Z_{{\varphi}^n}$ are $m$ and $\nu$ independent in the MS scheme and contain only poles with respect to $\eps$ variable. They are connected with the UV divergences of the equal time ($t_1=...=t_n$) diagrams at zero external momenta. In the framework of $\eps$ - regularization that is similar to the dimensional regularization so we can explore an analytical continuation to the $\eps<0$ region. In this region it is possible to set $m=0$ in the Eq.(\ref{cor1}). This yields

\begin{eqnarray}
\label{DV}
\nonumber {\bf D_v}({\bf x}) = D_0\frac{2^{-\eps}\Gamma(-\eps/2)}{{(4\pi)}^{d/2}\Gamma(d/2 + \eps/2)(d + \eps)}{|{\bf x}|}^{\eps}\times\\
\times\left[(d - 1 + \alpha + \eps){\delta}_{ij} + \eps(\alpha - 1)\frac{{\bf x}_i{\bf x}_j}{{\bf x}^2}\right],\quad\eps<0.
\end{eqnarray}

An arbitrariness of the $D_{0}$ parameter corresponds to the finite renormalization that does not affect the critical exponents. Let's then choose this parameter equal to

\begin{eqnarray}
\label{D0} D_0 = \frac{\Gamma(d/2 + \eps/2)(d + \eps)2^{\eps}}{\Gamma(1 - \eps/2)\Gamma(d/2)d},
\end{eqnarray}

\noindent so that the first fraction in the Eq.(\ref{DV}) contains a poles in $\eps$ only. It is convenient to divide then the correlator discussed into three parts with different $\eps$ - dependence:

\begin{eqnarray}
\label{D1}
{\bf D_v}({\bf x}) = {\bf D}({\bf x}) + {\bf \Delta}({\bf x}) + {\bf \bar \Delta }({\bf x}),\\
\label{D2}
{\bf D}({\bf x}) = -\frac{2 |{\bf x}|^{\eps}}{{d(4\pi)}^{d/2}\Gamma(d/2)}\left({\delta}_{ij} + (\alpha - 1)\frac{{\bf x}_i{\bf x}_j}{{\bf x}^2}\right),\\
\label{delti} { \bf \Delta}({\bf x}) = \frac{2(d - 1 + \alpha)}{{d(4\pi)}^{d/2}\Gamma(d/2)}\frac{1 - |{\bf x}|^{\eps}}{\eps},\quad {\bf \bar\Delta} = -\frac{2(d - 1 + \alpha)}{{d(4\pi)}^{d/2}\Gamma(d/2)}\frac{1}{\eps}.
\end{eqnarray}

Article \cite{TMF} was devoted to the modification of the steepest descent approach for the large-order asymptotes investigation of renormalization constants at the example of the well-known static $\varphi ^{4}$ model in MS scheme and dimensional regularization. According to this article UV divergences have a sense as the perturbation objects and must be treated as the fore-exponential factors. So the factor $\exp(S_{div})$ where $S_{div}$ absorbs all divergent terms of the action has to be expanded in McLoran series in $ \eps $ before the application of the steepest descent approach. As the expressions (\ref{delti}) contain the UV divergences (at $\eps\to 0$, ${\bf x}\to 0$) one need to include the corresponding terms to the divergent part of the action $S_{div}$ and consider them as fore-exponential factors. In Section VI we will see that these terms affect only an amplitude of the asymptotic investigated.

For $ \varphi ^{2} $ operator we can write in the Lagrangian variables with help of Exprs.(\ref{lag},\ref{1.5a},\ref{1.8})

\begin{eqnarray}
\label{1}\nonumber
\langle{\varphi}^2({\bf x}_0,t_0)\varphi'({\bf x}_1,t)\varphi'({\bf x}_2,t)\rangle = M^2\int D{\bf v}\int\limits_{{\bf c}_1(t)={\bf x}_1,{\bf c}_2(t)={\bf x}_2} ^{{\bf c}_1(t_0)={\bf c}_2(t_0)={\bf x}_0} D{\bf c}_1D{\bf c}_2\times\\
\times D{\bf c}_1^{\prime}D{\bf c}_2^{\prime} \exp\Bigg(-\frac 12{\bf v}D_{v}^{-1}{\bf v} - \nu Z_{\nu}{\bf c}_1'^2 - \nu Z_{\nu}{\bf c}_2'^2 +i{\bf c}_1'\dot {\bf c}_1+i{\bf c}_2'\dot {\bf c}_2 + \nonumber\\
+ ig{\bf c}'_1{\bf v}({\bf c}_1)+ig{\bf c}'_2{\bf v}({\bf c}_2)\Bigg)\Bigg [\int D{\bf v}\exp(-\frac{1}{2}{\bf v}D_{v}^{-1}{\bf v})\Bigg ]^{-1},
\end{eqnarray}
\begin{eqnarray}
\label{3}
M \equiv \frac{1}{(4\pi\nu T)^{d/2}}\Bigg(\int\limits_{{\bf c}(t)=0}^{{\bf c}(t_0)=0}D{\bf c}D{\bf c}' \exp(-\nu {\bf c}'^2 + i{\bf c}' \dot {\bf c})\Bigg)^{-1}.
\end{eqnarray}

Integration of the Expr.(\ref{1}) in ${\bf v}$ field yields

\begin{eqnarray}
\label{2}
M^2\int\limits_{{\bf c}_1(t)={\bf x}_1,{\bf c}_2(t)={\bf x}_2}^{{\bf c} _1(t_0)={\bf c}_2(t_0)={\bf x}_0} D{\bf c}_1D{\bf c}_2D{\bf c}_1' D{\bf c}_2' \exp\Bigg(-\nu Z_{\nu}({\bf c}_1'^2+{\bf c}_2'^2)+ \nonumber\\
+ i{\bf c}_1'\dot {\bf c}_1 + i{\bf c}_2' \dot {\bf c}_2 - u{\bf c}_1' \Bigg [{\bf D}({\bf c}_1 - {\bf c}_2) + {\bf \Delta}({\bf c}_1 - {\bf c}_2) + {\bf \bar\Delta}\Bigg ]{\bf c}_2'.
\end{eqnarray}

\noindent It's worth while to scale $u$ by $\nu$ to obtain a new fully dimensionless coupling constant $u$. After the expanding of the factor $S_{div}$ with UV divergent part of action

$$S_{div}=-u\nu {\bf c}_1^{\prime}{\bf \Delta}({\bf c}_1-{\bf c} _2){\bf c}_2^{\prime}-u\nu{\bf c}_{1}'{\bf \bar \Delta }({\bf c}_1-{\bf c}_2){\bf c}_{2}'-\nu (Z_{\nu} -1)({\bf c}_1^{\prime 2}+{\bf c}_2^{\prime 2})$$ the Expr.(\ref{2}) has a form

\begin{eqnarray}
\label{5}
M^2\int\limits_{{\bf c}_1(t)={\bf x}_1,{\bf c}_2(t)={\bf x}_2}^{{\bf c} _1(t_0)={\bf c}_2(t_0)={\bf x}_0}D{\bf c}_1D{\bf c}_2D{\bf c}_1^{\prime} D{\bf c}_2^{\prime}A^{[2]}B^{[2]}e^{-S^{[2]}},
\end{eqnarray}

\noindent where

\begin{eqnarray}
S^{[2]} = \nu({\bf c}_1'^2+{\bf c}_2'^2) - i{\bf c}_1' \dot {\bf c}_1 - i{\bf c}_2' \dot{\bf c}_2 + u\nu{\bf c}_1' {\bf D}({\bf c}_1 - {\bf c}_2) {\bf c}_2',
\end{eqnarray}
\begin{eqnarray}
\label{A}
A^{[2]} = \sum _{j=0}^{\infty}\frac{(-u\nu {\bf c}_1' {\bf \Delta}({\bf c}_1 - {\bf c} _2){\bf c}_2')^j}{j!},
\end{eqnarray}
\begin{eqnarray}
\label{B}
B^{[2]} = \sum _{j=0}^{\infty}\frac {(-u\nu{\bf c}_1'{\bf \bar\Delta}{\bf c}_2' - \nu(Z_{\nu} - 1)({\bf c}_1'^ 2 + {\bf c}_2'^2))^j}{j!}.
\end{eqnarray}

The similar expression can be written for $\langle{\varphi}^n({\bf x}_0,t_0)\varphi' ({\bf x}_1,t)\ldots\varphi' ({\bf x}_n,t)\rangle$ Green function. It can be represented via the path integral in ${\bf c}_i, {\bf c}'_i$  fields ($i=1,...,n$) with the action

\begin{eqnarray}
\label{actn}
S^{[n]} = \nu\sum\limits_{i=1}^n {\bf c}_i'^2 - i\sum\limits_{i=1}^n{\bf c}_i' \dot {\bf c}_i + \frac{u\nu}{2}\sum\limits_{i\neq l}{\bf c}_i'{\bf D}({\bf c}_i - {\bf c}_l){\bf c}_l'.
\end{eqnarray}

In the last term the vectorial indexes of the fields and the tensor ${\bf D}$ are implied to be contracted. The normalization factor equals to $M^{n}$ now. The boundary conditions are ${\bf c}_i(t_{0})={\bf x}_0$, ${\bf c}_i(t)= {\bf x}_i$. The
pre-exponential factors $A^{[n]}$ and $B^{[n]}$ are similar to (\ref{A},\ref{B}) with additional sums over $i$ index of the fields ${\bf c}_i, {\bf c}_i'$. Note that ${\bf c}_i, {\bf c}_i'$ fields play a role of generalized coordinates and momenta for the action $S^{[n]}$. So we can represent our system with a set of $n$ moving quasi-particles.

Now let's discuss the connection between the Green functions considered and the critical exponents. The singularities of the Green function and renormalization constants of the latter can be related via the identity (\ref{2.11}). The renormalization constant $Z_{{\varphi}^n}$ determines uniquely the anomalous dimension ${\gamma}_{{\varphi}^n}$.

The l.h.s. of the Eq.(\ref{2.11}) is free of singularities and so does the right one: the singularities of $Z_{{\varphi}^n}$ are canceled by those of non-renormalized function $\langle{\varphi}^n\varphi' \ldots\varphi' \rangle$. We could use this to compute the singularities of $Z_{{\varphi}^n}^{(N)}$ by stating:

\begin{equation}
\label{str}
\Bigg(Z_{{\varphi}^n}^{(N)}\langle{\varphi}^n\varphi' \ldots\varphi' {\rangle}^{(0)} + Z_{{\varphi}^n}^{(N-1)}\langle{\varphi}^n\varphi' \ldots\varphi' {\rangle}^{(1)} + \ldots + Z_{{\varphi}^n}^{(0)}\langle{\varphi}^n\varphi' \ldots\varphi' {\rangle}^{(N)}\Bigg ) = \mbox{finite}.
\end{equation}

\noindent Henceforth $X^{(N)}$ denotes the $N$-th order of the expansion for the value $X$ in $u$. Usually expansions in the quantum-field theories are asymptotic ones with an exponential growth of the coefficients. Such growth would make all terms in the Eq.(\ref{str}) except the first and the last ones irrelevant at $N\to\infty$. Then Eq.(\ref{str}) could be easily resolved for $Z^{(N)}$ coefficients. The existence of the finite radius of convergence will be shown for Kraichnan model. Then a large set of terms in (\ref{str}) become relevant and we face the problem of the unknown terms $\langle{\varphi}^n\varphi' \ldots\varphi' {\rangle}^{(i)}, i<N$: we cannot pick out all the singularities present in the Eq. (\ref{str}).

That's why we shall use the identity $$\ln\langle{\varphi}^n\varphi' \ldots\varphi' {\rangle}_R = -\ln Z_{{\varphi}^n} + \ln\langle{\varphi}^n\varphi' \ldots\varphi' \rangle $$ The l.h.s of the expression is finite again at $\eps\to 0$. To restore the N-th coefficient of the expansion for $\ln Z_{{\varphi}^n}$ in $u$ we will investigate the residue in $\eps$ for McLoran expansion of $\ln\langle{\varphi}^n\varphi' \ldots\varphi' \rangle$. The "replica trick" that is based on the identity $$ \ln\langle{\varphi}^n\varphi' \ldots\varphi'\rangle = \lim_{L\to0}\frac{\partial}{\partial_{L}}\langle{\varphi}^n\varphi' \ldots\varphi'\rangle^{L}. $$ will be used to treat $\ln\langle{\varphi}^n\varphi' \ldots\varphi' \rangle$. The Green function in $L$ - power in the last expression is then substituted by the path integral representation of $\langle{\varphi}^n\varphi' \ldots\varphi' \rangle$ with all integration variables considered now as $L$ - dimensional ones.

\section{Instanton analysis and immovable particles.}

As in classical works \cite{Lipatov,Zinn} we add an integration $\oint du/u^{N+1}$ to produce the $N$ - th order of the perturbation expansion. Summarizing all the remarks mentioned above the $N$ - th order term of the expansion of $\ln Z_{\varphi ^{n}}$ in $u$ can be written as

\begin{eqnarray}
\label{base}\nonumber
(\ln\langle{\varphi}^n\varphi' \ldots\varphi' \rangle)^{(N)} = \frac{1}{2\pi i}\lim_{L\to0}\frac{\partial}{\partial L}\mathop{\mathrm{residue}}\limits_{\eps = 0}\oint\frac{du}{u^{N+1}}\prod_{\alpha=1}^{L}\int d T_{\alpha}\int d{\bf x}_{0\alpha}\times\\
\nonumber
\times\int d({\bf x}_{2\alpha} - {\bf x}_{1\alpha})\int d({\bf x}_{3\alpha} - {\bf x}_{1\alpha})\ldots \int d({\bf x}_{n\alpha} - {\bf x}_{1\alpha}) M^{n}_\alpha\times\\
\times\int\limits_{\{{\bf c}_{i\alpha}(t_{\alpha}) = {\bf x}_{i\alpha}\}}^{\{{\bf c}_{i\alpha}(t_{0\alpha}) = {\bf x}_{0\alpha}\}}\Bigg (\prod_{i=1}^{n} D{\bf c}_{i\alpha }D{\bf c}_{i\alpha}'\Bigg ) {\bf A}^{[n]}_\alpha {\bf B}^{[n]}_\alpha\exp (-S^{[n]}_{\alpha} - i{\bf k}({\bf x}_{2\alpha} - {\bf x}_{1\alpha}))
\end{eqnarray}

\noindent with the action (\ref{actn}). The variables ${T}, {\bf x}_i, {\bf c}_i, {\bf c}'_i$ have now an additional replica index $\alpha =1...L$.

Let us shift the variables to eliminate all dependencies of the path integral limits in coordinates

\begin{eqnarray}
\label{shift}
{\bf c}_{i\alpha}(\tau_{\alpha}) = \bar{\bf c}_{i\alpha}(\tau_{\alpha}) + {\bf x}_{i\alpha} - \frac{({\bf x}_{i\alpha} - {\bf x}_{0\alpha})(\tau_{\alpha} - t_{\alpha})}{T_{\alpha}},
\end{eqnarray}

\noindent the new fields $\bar {\bf c}_{i\alpha}$ have zero boundary conditions. Sometimes we will return to the original notation ${\bf c}_i$ and we will omit the index $\alpha$ for brevity.

The steepest descent approach must be applied to the expression (\ref{base}) with respect to all integrations except the one over the scale parameter of the model that has an essential non-instanton form. Indeed, overall UV divergences of interest arise from the integration over the scale parameter and can be extracted with the help of the integration by parts. We will show that at the instanton solution the value $y=|{\bf x_{2\alpha}} - {\bf x_{1\alpha}}|$ doesn't depend actually on $\alpha$ and turn out to be the scale parameter.

We proceed now to the finding of the saddle-point of the action in (\ref{base}). Let us scale the variables to figure out the $N$ dependence of the action. The renormalization constants considered are independent on renormalized diffusivity $\nu$ and ${\bf k}$ momentum so they can be also scaled:

\begin{eqnarray}
\label{tjazh}
{\bf c}_i' \to N{\bf c}_i' \quad {\bf k}\to N{\bf k},\quad\nu=\eta/N
\end{eqnarray}

\noindent (we perform this scaling both for the Expr. (\ref{base}) and for the factor $M$ integral representation (\ref{3}); the Jacobians arising are constant and cancel mutually). Such a scaling produces $N^{ndL/2}$ factor for the expression considered. The value $S^{[n]}_{\alpha}$ has now the following form:

\begin{eqnarray}
\nonumber S^{[n]}_{\alpha} = N\left\{\eta\sum\limits_{i=1}^n {\bf c}_i'^2 - i\sum\limits_{i=1}^n{\bf c}_i' \dot {\bf c}_i +
\frac{u\eta}{2}\sum\limits_{i\neq j}{\bf c}_i' D({\bf c}_i - {\bf c}_j){\bf c}_j'\right\} \equiv N\widetilde{S}^{[n]}_\alpha,
\end{eqnarray}

\noindent all the fields are assumed to have the replica index $\alpha$. In the framework of the instanton approach we vary the functional $$S\equiv N\sum_{\alpha}\Bigg (\widetilde{S}^{[n]}_{\alpha} + i{\bf k}({\bf x}_{2\alpha} - {\bf x}_{1\alpha})\Bigg ) + N\ln u$$ with respect to all the variables. The variations in $\bar{\bf c}_m, {\bf c}_m'$ yield

\begin{eqnarray}
\label{I}
\frac{\delta {S}}{\delta \bar{\bf c}_m} = 0\quad\hfill\Rightarrow\quad\qquad -i\dot{\bf c}_m' = u\eta\sum_{{}^{\hskip 0.2cm l}_{l\neq m}} {\bf c}_m' \frac{\partial D({\bf c}_m - {\bf c}_l)}{\partial ({\bf c}_m - {\bf c}_l)}{\bf c}_l',\\
\label{II}
\frac{\delta S}{\delta {\bf c}_m^{\prime}} = 0\quad\hfill\Rightarrow\qquad i\dot{\bf c}_m = u\eta\sum_{{}^{\hskip 0.2cm l}_{l\neq m}}D({\bf c}_m - {\bf c}_l){\bf c}_l' + 2\eta{\bf c}_m'.
\end{eqnarray}

The dependence on coordinates ${\bf x}_i$ is represented in the functional $S$ only by the fields ${\bf c}_i$, ${\bf c}_j$ (see the Expr. (\ref{shift})). So the variation in ${\bf x}_0$ yields the full momentum conservation law for the quasi-particles:

\begin{eqnarray*}
\label{III} \frac{\delta {S}}{\delta {\bf x}_0} =
0\quad\quad\Rightarrow\quad\quad \sum\limits_{l=1}^n {\bf c}_l' = 0.
\end{eqnarray*}

The variations with respect to the coordinates ${\bf x}_m - {\bf x}_1$ ($m\geq 3$) yield:

\begin{eqnarray}
\label{IV} \frac{i}{T}\int\limits_t^{t_0}d\tau {\bf c}_m' + u\eta\int\limits_t^{t_0}d\tau\frac{t_0 - \tau}{T}\sum_{{}^{\hskip 0.2cm l}_{l\neq m}} {\bf c}_m' \frac{\partial D({\bf c}_m - {\bf c}_l)}{\partial {(\bf c}_m - {\bf c}_l)} {\bf c}_l' = 0, \quad m\geq 3.
\end{eqnarray}

\noindent Combining together Eq.(\ref{I}) and Eq.(\ref{IV}) we get the equation

\begin{eqnarray}
\label{pn-boundary}
\frac{1}{T}\int\limits_t^{t_0} d\tau {\bf c}_m' - \frac{1}{T}\int\limits_t^{t_0} d\tau(t_0 - \tau)\dot{\bf c}_m' = {\bf c}_m'(t)=0.
\end{eqnarray}

The last expression can be considered as a boundary condition at $\tau=t$ on the field ${\bf c}_m' (\tau)$ ($m\geq3$). The Eq.(\ref{I}) is a first order differential equation with respect to ${\bf c}_m'$ and has a zero boundary condition (\ref{pn-boundary}). It has then the trivial solution

\begin{eqnarray}
\label{trivial1} {\bf c}_m^{\prime}(\tau)=0,\qquad t\leq\tau\leq t_0 \qquad m\geq 3, \quad \alpha =1\ldots L
\end{eqnarray}

that is locally unique. The Eq.(\ref{II}) provides:

\begin{eqnarray}
\label{trivial2}
\dot{\bf c}_{m} = 0\qquad {\bf c}_m(\tau) = {\bf x}_m = {\bf x}_0\qquad m\geq 3, \quad\alpha = 1\ldots L.
\end{eqnarray}

This means that the instanton sought for implies only a couple of quasi-particles in motion while the others rest at the point ${\bf x}_0$. The saddle-point equations reduce then to the $n=2$ case with only four nontrivial fields $\bar{\bf c}_1, \bar{\bf c}_2, {\bf c}_1', {\bf c}_2'$ (replica index $\alpha$ is assumed).
Such a skewness of the instanton is explained by the term ${\bf k}({\bf x}_{2\alpha} - {\bf x}_{1\alpha})$ of the variable action. Indeed, as we are interested in renormalization constants that don't depend on a conjugate momentum running through diagrams of the response function considered we have set all momenta except ${\bf k}$ in the Expr. (\ref{base}) equal to zero. The momentum ${\bf k}$ can't be equal to zero since the object considered would turn out to be divergent. Note that this divergence is unrelated to the problem as it has a trivial power form and doesn't affect on the radius of convergence analyzed. Nevertheless the accurate treatment requires ${\bf k}\neq 0$ that causes the skewness obtained.

Thus the analysis of the stationarity equations demonstrates the trivial solution (\ref{trivial1},\ref{trivial2}) for the variables ${\bf x}_k$, $k\ge 3$. The problem reduces then to the $\varphi^2$ case and we can limit ourselves to the study of the saddle-point of the following nontrivial integral representation

\begin{eqnarray}
\label{p2-main}
\frac{1}{2\pi i}\oint\frac{du}{u^{N+1}}\prod_{\alpha}\int d{\bf x}_0\int d{\bf x}\int\limits_0^{\infty} dT \int D\bar{\bf c}_1D\bar{\bf c}_2 D{\bf c}_1' D{\bf c}_2' A^{[2]}B^{[2]}M^2 e^{-S_\alpha^{[2]}},
\end{eqnarray}
that represents the case $n=2$. Indeed, the substitution of (\ref{I},\ref{II}) to the expressions for $A^{[n]}_\alpha, B^{[n]}_\alpha$ reduces their values to the two-particle case $A^{[2]}, B^{[2]}$ and $S^{[n]}_{\alpha}$ as well.

Let us introduce new variables: ${\bf p} = {\bf c}_1' - {\bf c}_2', {\bf P} = {\bf c}_1' + {\bf c}_2', \bar{\bf q} = \bar{\bf c}_1 - \bar{\bf c}_2, \bar{\bf Q} = \bar{\bf c}_1 + \bar{\bf c}_2$. The fields $\bar{\bf q}, \bar{\bf Q}$ satisfy zero boundary conditions. Sometimes we will use the notation ${\bf q} = {\bf c}_1 - {\bf c}_2, {\bf Q} = {\bf c}_1 + {\bf c}_2$ for brevity. The variables introduced correspond to the center mass frame of reference of the quasi-particles. The expression (\ref{p2-main}) transforms into:

\begin{eqnarray}
\frac{1}{2\pi i}\oint\frac{du}{u^{N+1}}\int d{\bf x}_0\int d{\bf x}\int\limits_0^{\infty} dT M^2A^{[2]}B^{[2]}\int D{\bf p}D\bar{\bf q}D{\bf P}D\bar{\bf Q} e^{-N\widetilde{S}},\\
\label{swaw}
\widetilde{S} = \sum_{\alpha=1}^L(\widetilde{S}_\alpha^{[2]} + i{\bf k}{\bf x}_\alpha) + \ln u, \qquad {\bf x} = {\bf x}_2 - {\bf x}_1, \\
\label{p2-main-general}
\widetilde{S}^{[2]}_\alpha = -\frac{i}{2}{\bf p}\Bigg (\dot{\bar{\bf q}} + \frac{{\bf x}}{T}\Bigg ) - \frac{i}{2}{\bf P}\Bigg (\dot{\bar {\bf Q}} + \frac{2{\bf x}_0 - {\bf x}_1 - {\bf x}_2}{T}\Bigg) + \frac{u\eta}{4}({\bf P} + {\bf p})D({\bf q})({\bf P} - {\bf p})
\end{eqnarray}

\noindent All the variables in the r.h.s. of the Expr.(\ref{p2-main-general}) are assumed to have the replica index $\alpha$. Zero conditions on variations of the action $\widetilde{S}$ in $\bar{\bf q}, \bar{\bf Q},{\bf p}, {\bf P}, {\bf x}, {\bf x}_0$ and $u$ give the instanton equations.

It is convenient to carry out the calculation in two steps. First we solve the instanton equations in fields ${\bf P}, \bar{\bf Q}, {\bf p}, \bar{\bf q}$ and the variable ${\bf x}_0$. We use their solution to simplify the functional $\widetilde{S}$. Then we vary the simplified functional $\widetilde{S}$ in ${\bf x}_\alpha$ and $u$ we obtain saddle-point equations in a simplified manner and we turn out to be successful in solving them.

\section{Two quasi-particle saddle-point solution}

The saddle-point of the action $\widetilde{S}$ with respect to ${\bf x}_0, \bar {\bf Q}, {\bf P}, \bar {\bf q}, {\bf p}$ is determined by the following equations:

\begin{eqnarray*}
\frac{\delta \widetilde{S}}{\delta{\bf x}_0} = 0\quad\hfill\Rightarrow\quad\quad\quad\quad\quad\quad\quad\quad\quad\quad\quad\quad\quad\quad\quad \dot{\bf P} = 0,\\
\frac{\delta \widetilde{S}}{\delta \bar{\bf Q}} = 0\quad\hfill\Rightarrow\quad\quad\quad\quad\quad\quad\quad\quad\quad\quad\quad\quad\quad\quad\quad {\bf P} = 0,\\
\frac{\delta \widetilde{S}}{\delta{\bf P}} = 0\quad\hfill\Rightarrow\quad\quad\quad\quad\quad\quad\quad\quad i\dot{\bf Q} = u\eta {\bf D}({\bf q}){\bf P} + 2\eta {\bf P},\\
\frac{\delta \widetilde{S}}{\delta \bar {\bf q}} = 0\quad\hfill\Rightarrow\quad\quad\quad i\dot{\bf p} = \frac{u\eta}{2}\frac{\partial }{\partial {\bf q}} \Bigg[{\bf p}{\bf D}({\bf q}){\bf p} - {\bf P}{\bf D}({\bf q}){\bf P}\Bigg ],\\
\frac{\delta \widetilde{S}}{\delta{\bf p}} = 0\quad\hfill\Rightarrow\quad\quad \quad\quad\quad\quad\quad\quad i\dot{\bf q} = 2\eta {\bf p} -u\eta {\bf D}({\bf q}){\bf p}
\end{eqnarray*}

\noindent with the following boundary conditions

\begin{eqnarray}
\nonumber
{\bf q}(t) = -{\bf x},\quad\quad{\bf Q}(t) = {\bf x}_1 + {\bf x}_2,\\
\nonumber
{\bf q}(t_0) = 0,\quad\quad\quad\quad {\bf Q}(t_0) = 2{\bf x}_0.
\end{eqnarray}

\noindent To find the solution we suppose that the vector ${\bf p}$ is parallel to ${\bf q}$. In fact one can show that this is the only solution of the original system satisfying the given boundary conditions. Similar to \cite{Chertkov} the simplified system is trivially integrated:

\begin{eqnarray}
\label{sol1}
{\bf Q}(\tau) = 2{\bf x}_0 = {\bf x}_1 + {\bf x}_2,\qquad {\bf P}(\tau) = 0,\\
\label{sol2}
{\bf p}(\tau) = \frac{i\dot {\bf q}(\tau)}{2\eta - u\eta D({\bf q}(\tau))},\qquad\dot{\bf q}(\tau) = \frac{I_1({\bf x})}{T}\sqrt{2\eta - u\eta D({\bf q}(\tau))},\\
I_1({\bf x}) = \int_0^x \frac{dz}{\sqrt{2\eta - u\eta D(z)}},\qquad{D}(z)\equiv-\frac{2\alpha |z|^{\eps}}{{d(4\pi)}^{d/2}\Gamma(d/2)},
\end{eqnarray}

\noindent where $D(z)$ and $x$ is the projections of the tensor ${\bf D}({\bf q})$ and of the vector ${\bf x}$ to the direction of the vector ${\bf p}$. As usual the index $\alpha$ is implied for the time variables $T, \tau$, all fields and the variables $x, {\bf x}$.

The functional $\widetilde{S}$ at the saddle-point equals

\begin{eqnarray}
\label{sstar}
\widetilde{S} = \sum_{\alpha=1}^L(\widetilde{S}^{[2]}_\alpha - i{\bf k}{\bf x}_\alpha) + \ln u,\qquad \widetilde{S}^{[2]}_\alpha = \frac{I_{1}^2({\bf x}_\alpha)}{4T_\alpha},
\end{eqnarray}

\noindent where the replica index is indicated explicitly.

Let's note that due to $P=0$ the following identity holds ${\bf c}_1' = -{\bf c}_2'$ and the $\bar\Delta$ - term in $B^{[2]}$ factor (\ref{5}) is canceled by the renormalization $(Z_{\nu}-1)$ of the diffusivity $\nu$ and thus $B^{[2]}\equiv1$ (\ref{B},\ref{delti},\ref{Z}). The pre-exponential factor $A^{[2]}$ in the saddle point is given by

\begin{eqnarray}
\label{A22}
A^{[2]}_\alpha = \sum_{j=0}^\infty\frac{1}{j!} \Bigg(-\frac{u\eta NI_1({\bf x}_\alpha)I_\Delta({\bf x}_\alpha)}{4T_\alpha}\Bigg )^j,\qquad I_\Delta({\bf x}) = \int_0^x\frac{\Delta (z)dz}{\sqrt{2\eta - u\eta D(z)}^3}.
\end{eqnarray}

\section{A saddle-point solution over $u$ and ${\bf x}_\alpha$}

It is obvious that the variations in $u$ and ${\bf x}_\alpha$ of $\widetilde{S}$ given by the Expr. (\ref{swaw}) and $\widetilde{S}$ given by the Expr. (\ref{sstar}) are explicitly the same. The form (\ref{sstar}) is much more convenient: the variation with respect to $u, {\bf x}_{\alpha}$

\begin{eqnarray}
\label{varu}
\frac{\delta\widetilde{S}}{\delta u} = 0\quad\Rightarrow\quad \sum_{\alpha=1}^L \frac{u\eta I_1({\bf x}_\alpha)I_2({\bf x}_\alpha)}{4T_\alpha} = -1,\\
\nonumber
\quad I_2({\bf x}_\alpha) = \int_0^{x_\alpha}\frac{dz D(z)}{[\sqrt{2\eta - uD(z)}]^3},\\
\label{varx}
\frac{\delta \widetilde{S}}{\delta {\bf x}_\alpha} = 0\quad\Rightarrow\quad \frac{I_1({\bf x}_\alpha)}{2T_\alpha\sqrt{2\eta - uD({\bf x}_\alpha)}} = i{\bf k}.
\end{eqnarray}

These equations include $T_\alpha$ parameter: let's suppose their solution exist. Substituting it into the expression (\ref{sstar}) we obtain the action $\widetilde{S}^{[2]}_\alpha = NI_1^2/T_\alpha$ i.e. the integration over $T_\alpha$ has the form $$\int_0^\infty dT_\alpha f(T_\alpha)\exp{\Bigg(-\frac{NI_1^2}{T_\alpha} \Bigg )}.$$ The convergence of this integral is assured by $f(T)$ factor that absorbs all the fluctuation integrals over the fields and $u, {\bf x}_0, {\bf x}$ variables. It is easy to see that the main contribution to the integration over $T_\alpha$ at $N\rightarrow\infty$ is determined by large $T_\alpha$ region. Let's return to the Eq.(\ref{varu}),(\ref{varx}) analysis. The l.h.s. of the Eq.(\ref{varx}) rests non-zero in the $T_{\alpha}\to\infty$ limit only if the value $\sqrt{2\eta-uD({\bf x}_\alpha)}$ is small ($\sim T_\alpha^{-1}$). On the other hand the Eq.(\ref{varu}) is fulfilled only for large values of $I_2$, i.e as the denominator $\sqrt{2\eta-uD(z)}$  reduces to zero at the point that tends to $x+0$ at large $T_\alpha$. Therefore the solution of the system (\ref{varu}), (\ref{varx}) can be written at large $T_\alpha$ as

\begin{eqnarray}
\label{xu} x_\alpha = y + \frac{\delta_\alpha}{T_\alpha^2},\qquad u = \frac{2}{D(y + \delta_0\sum_\alpha T_\alpha^{-2})}
\end{eqnarray}
with $\delta_0, \delta_\alpha$ being some constants and $y$ being an arbitrary scale parameter. The substitution of this solution into the Eqs.(\ref{varu}), (\ref{varx}) allows to calculate $\delta_0, \delta_\alpha$ constants. In the limit $T\to\infty$ this saddle-point expressed in ${\bf x}_\alpha, u$ is $${\bf x}_\alpha = y\frac{\bf k}{|{\bf k}|},\quad\alpha=1\ldots L,\qquad u = \frac{2}{D(y)}.$$ The arbitrary scale parameter appearance is the usual fact at instanton investigation of scale invariant models \cite{Lipatov}. Its correct consideration demands an insertion of a unit decomposition in the initial expression (\ref{base}) $$1\equiv\int_0^\infty dy \delta (y-|{\bf x}_{\alpha=1}|).$$ The integration over the scale parameter can be made explicit by Faddeev-Popov method. It turns out to have the non-saddle-point form and its singularities at $\eps\to 0$ determine the renormalization constants and the critical exponents (the same situation is observed for $\varphi^4$ model \cite{TMF}). The $\delta$ - function mentioned above plays an important role as it solves the so called "zero mode" problem \cite{Zinn} in the carrying out of the integration over the fluctuations. Besides it produces an additional $\sqrt{N}$ factor as a result of the elimination of the integration over one of the degrees of freedom ("zero mode") in the numerator of the fluctuation integral. The other factor $N^{-(ndL+1)/2}$ appears from the fluctuation integration over the variables $u, {\bf x}_0, {\bf x}_{i\alpha}$ ($i=2\ldots n, \alpha=1\ldots L$). Together with $N^{ndL/2}$ factor appeared from the scaling (\ref{tjazh}) the power function in $N$ reduces to unit.

The last question we have to discuss in this Section is the saddle-point method applicability. It should be mentioned that the fluctuation integration can be carried out at $\eps=0$. Though the longitudinal projector $$P_{ab}^{||}=\frac{ ({\bf c}_i - {\bf c}_j)_a({\bf c}_i - {\bf c}_j)_b}{({\bf c}_i - {\bf c}_j)^2}$$ in the correlator (\ref{cor1},\ref{DV}) cannot be expanded in the series of the fluctuations of immovable quasi-particles variables ($i,j>2$) with respect to the saddle-point, the integration over these fluctuations ($\delta{\bf c}_i, \delta{\bf c}_i', i>2$) is Gaussian and can be correctly performed by the means of transition from the path integral in $\delta{\bf c}_i'$ ($i>2$) to the integration over $\delta{\bf c}_{i||}', \delta{\bf c}_{i\perp}'$ fields that represent the longitudinal and transversal projections of $\delta {\bf c}_i'$ field on the $\delta{\bf c}_i$ direction.

The consistent analysis of the higher variations of the action (\ref{actn}) around the saddle point demands their examination in the large $T$ limit as it is this region that contributes to the integration over $T$ of the expression studied. Using the explicit solution (\ref{sol1},\ref{sol2}) one can show that the higher variations of the action considered rest finite in the large $T_\alpha$ limit. This assures the consistency of the instanton approach and the existence of the fluctuation integral.

\section{The integration over the scale parameter}

Finally the result of instanton approach is

\begin{eqnarray}
\label{p2-main-x-2}
\ln Z_{\varphi ^{n}}^{(N)}\sim \lim_{L\to 0}\frac{\partial}{\partial L}\mathop{\mathrm{residue}}_{\eps=0}\left\{K^L \int_0^\infty\frac{dy}{y}\Bigg(\frac{D(y)}{2}\Bigg)^N\kappa(y)\prod_{\alpha=1}^LA^{[2]}_\alpha \right\}.
\end{eqnarray}

The factor $K^L$ arises from the $L$ - dimensional integration over $T$. The function $\kappa(y)$($\kappa(0)<\infty$) results from all the fluctuation integrations, besides the factor $\prod_\alpha e^{i|{\bf k}|y}$ is also included in $\kappa$. The factor $1/y$ is extracted from fluctuation integrations without explicit calculation by the means of the dimension analysis. Indeed, the value $\ln Z_{\varphi^{n}}$ is determined by logarithmic divergences of the diagrams corresponding to $\langle{\varphi}^2\varphi' \ldots\varphi' {\rangle}^{(N)}$ function. This gives the logarithmic behavior of the integral in $y$ (\ref{p2-main-x-2}). The convergence of the Expr. (\ref{p2-main-x-2}) at large $y$ is ensured by $\kappa(y)$ factor. Let's remind that this factor also depends on $L$.

Using the formula (\ref{xu}) we note that $I_\Delta ({\bf x}_\alpha)\sim T_\alpha$ at the saddle-point, so the Expr. (\ref{A22}) can be rewritten as $$A^{[2]} = \sum_{j=0}^\infty\frac{1}{j!}\Bigg( -\frac{K(\eps)N(1 - y^\eps f(\eps))}{\eps}\Bigg)^j$$ with $K(\eps)$ and $f(\eps)$ being regular functions in $\eps$, $K(0)\neq 0, f(0)=1$. The extraction of UV divergences (i.e. poles in $\eps$) have to be executed in the framework of the formalism developed in {\cite{TMF}}. The parameter $N\eps$ is supposed to be small. Using the explicit formula for $D(y)$ (\ref{sol2}) the divergences at $\eps\to 0$ can be easily extracted by integration by parts:

\begin{equation}
\label{p2-result}
\int_0^\infty\frac{dy}{y}y^{N\eps}\kappa(y)\frac{(1-y^\eps)^j}{\eps^j} = \frac{\kappa(0)}{\eps^{j+1}}\sum_{s=0}^{j}\frac{(-1)^{j-s}j!}{(N+s)s!(j-s)!}.
\end{equation}

Due to the choice of ${\bf D_v}$ correlator (see definition of $D_0$ constant by the Expr. (\ref{D0})) the studied object (\ref{p2-main-x-2}) doesn't contain $N\eps$ corrections. Then $A^{[2]}$ term doesn't affect the position and the type of the singularity and contributes at large order in $1/N$ only to the amplitude of the asymptotes. The final result for the single pole in $\eps$ of the renormalization constant series in $u$ is given then by the following formula

\begin{equation}
\label{inty}
\ln Z^{(N)}_{\varphi^n}\sim \left\{ -\frac{\alpha}{{(4\pi)}^{d/2}\Gamma(d/2)d}\right\}^N.
\end{equation}

\section{Anomalous dimensions resummation}

We proceed now to the critical exponents ${\Delta}_{{\varphi}^n}$ analysis. The general expression  for critical exponents is

\begin{eqnarray}
\nonumber {\Delta}_{{\varphi}^n} = d_{{\varphi}^n} + {\gamma}_{{\varphi}^n} = n(-1 + \eps) + {\gamma}_{{\varphi}^n},
\end{eqnarray}

\noindent where $d_{{\varphi}^n}$ and ${\gamma}_{{\varphi}^n}$ stand for the canonical and anomalous dimensions of the composite operator ${\varphi}^n$. Our main goal is to analyze the expansion of the anomalous dimension ${\gamma}_{{\varphi}^n}$ in $\eps$ using the information on the series singularities. These singularities are of the main interest since they can be used for resummation of the series. General theory \cite{Zinn} provides

\begin{eqnarray}
\label{gam}
{\gamma}_{{\varphi}^n} = D_{\mu}\ln Z_{{\varphi}^n} = \beta(u)\frac{\partial}{\partial u}\mathop{\mathrm{residue \hskip 0.1cm }}\limits_{\eps = 0}[\ln Z_{{\varphi}^n}]\Bigg |_{u=u_*} = -2 u\frac{\partial}{\partial u}\mathop{\mathrm{residue \hskip 0.1cm}}\limits_{\eps =0}[\ln Z_{{\varphi}^n}]\Bigg |_{u=u_*}
\end{eqnarray}

\noindent with $u_*$ being a fixed point. The latter is known exactly for Kraichnan model:

\begin{eqnarray}
\label{u*}
u_*=2\eps\frac{{(4\pi)}^{d/2}\Gamma(d/2)d}{(d-1+\alpha)}.
\end{eqnarray}

\noindent The substitution of the Expr. (\ref{inty}, \ref{u*}) to the formula (\ref{gam}) gives the transformation of $u$ - expansion of the anomalous dimension ${\gamma}_{{\varphi}^n}$ to the series in $\eps$ considered at the large order of perturbation theory:

\begin{eqnarray*}
{\gamma}_{\varphi^n}=\sum\limits_{N\ge 0}{\gamma}_{\varphi_n}^{(N)}\eps^N\\
{\gamma}_{{\varphi}^n}^{(N)}\sim{\left[\frac{-2\alpha}{(d-1+\alpha)}\right]}^N
\end{eqnarray*}

The absence of a power term $N^b$ in the resulting asymptote (i.e. $b=0$ in the Expr. (\ref{nomer})) indicates the simple pole singularity. The convergence radius $R_c$ and the position of the nearest singularity ${\eps}_c$ for the anomalous dimension expansion in $\eps$ is easily found to be: $$\eps_c = -\frac{d - 1+ \alpha}{2\alpha},\qquad R_c=|\eps_c|.$$ This information is used for resummation the ${\gamma}_{{\varphi}^n}$ series: we extract the singularity of the expansion adopting simple rational representation:

\begin{eqnarray}
\label{rx-reexp}
{\gamma}_{\varphi^n} = \sum\limits_{k=1}^{\infty}{\gamma}_{\varphi^n}^{(k)}\eps^k = \frac{\sum\limits_{k=1}^{\infty}\widetilde{\gamma}_{\varphi^n}^{(k)}\eps^k}{\eps-\eps_c}.
\end{eqnarray}

\noindent Term by term comparison of these expansions for ${\gamma}_{\varphi^n}$ yields  the following identities for the few first coefficients ${\gamma}_{{\varphi}^n}^{(k)}$ and $\widetilde{\gamma}_{\varphi^n}^{(k)}$

\begin{eqnarray}
\nonumber
\widetilde{\gamma}_{\varphi^n}^{(1)} = -{\gamma}_{\varphi^n}^{(1)}{\eps}_c,\quad\widetilde{\gamma}_{\varphi^n}^{(2)} = {\gamma}_{\varphi^n}^{(1)} - {\gamma}_{\varphi^n}^{(2)}\eps_c,
\end{eqnarray}

We now need the values of the coefficients ${\gamma}_{\varphi^n}^{(k)}$ to carry out the resummation. The anomalous dimension ${\gamma}_{{\varphi}^n}, n>1$ is calculated up to the second order in \cite{Adzhemyan} and the coefficients are given by the formulae:

\begin{eqnarray*}
{\gamma}_{\varphi^n}^{(1)} = -\frac{\alpha n(n-1)d}{d-1+\alpha},\\
{\gamma}_{\varphi^n}^{(2)} = 2\frac{\alpha(\alpha-1)n(n-1)(d-1)}{(d-1+\alpha)^2}+\frac{{\alpha}^2n(n-1)(n-2)dh(d)}{(d-1+\alpha)^2},\\
h(d) = \sum\limits_{k=0}^{\infty}\frac{k!}{4^k(1+d/2)\dots(k+d/2)} = F(1,1;1+d/2;1/4),
\end{eqnarray*}

\noindent where $F$ represents generalized hyper-geometric function. The resummation is then straightforward and gives the following coefficients $\widetilde{\gamma}_{\varphi^n}^{(k)}, k=1,2$:

\begin{eqnarray*}
\widetilde{\gamma}_{{\varphi}^n}^{(1)}=-n(n-1)\frac{d}{2},\qquad\widetilde{\gamma}_{{\varphi}^n}^{(2)}=-n(n-1)+\frac{\alpha(n-2)dh(d)}{2(d-1+\alpha)},\\
\end{eqnarray*}

\noindent and the anomalous dimensions equals to:

\begin{eqnarray}
\label{rx-res}
{\gamma}_{{\varphi}^n} = -\frac{2\alpha n(n-1)\eps} {d-1+\alpha(1+2\eps)}\left[\frac{d}{2}+\eps - \frac{\alpha d(n-2)h(d)}{2(d-1+\alpha)}\eps +O({\eps}^2)\right].
\end{eqnarray}

Fortunately, this result can be verified: for $n=2$ an exact value of the anomalous dimension is known:

\begin{eqnarray*}
{\gamma}_{{\varphi}^2} = -\frac{2\eps\alpha(d+2\eps)}{d-1+\alpha(1+2\eps)},
\end{eqnarray*}

while the resummation (\ref{rx-res}) for anomalous dimension ${\gamma}_{{\varphi}^2}, n=2$ yields

\begin{eqnarray*}
{\gamma}_{{\varphi}^2} = -\frac{2\alpha\eps\left[d+2\eps+O({\eps}^2)\right]}{d-1+\alpha(1+2\eps)}.
\end{eqnarray*}

We see that no high order corrections to the resummation result occur in this case; resummation gives the exact answer. The singularity found for $Z_{{\varphi}^2}$ turned out to be unique. This case illustrates the information about the type and the position of the singularity to allow us appreciably supplement the information obtained from direct perturbation expansion.

\section{Conclusions}

This article demonstrates that the instanton approach applies to the large-order analysis of the dynamical model. We can state that the study of the behavior of the convergent series is a more difficult problem than that for the divergent ones. Using the instanton approach one can state that the perturbation series for the scaling dimensions of the operators considered are convergent with the finite convergence radius calculated. In each case taking into account the information about singularity type of the series permit to accelerate considerably the convergence of these series.

Concerning the saturation problem \cite{Chertkov} one can say that our results don't prove its existence. It is possible that two orders of the $\eps$ expansion are not sufficient to observe the saturation. Another possibility is that there is no saturation in the studied model (saturation was predicted for the Kraichnan model with the transversal velocity field $\alpha =0$ \cite{Chertkov}, where our results can are not valid).

\end{document}